\documentclass[twocolumn,showpacs,preprintnumbers,amsmath,amssymb,aps,prb,10pt]{revtex4-1}

\usepackage{graphicx}
\usepackage{subfigure}
\usepackage{dcolumn}
\usepackage{bm}

\newcommand{\beq}{\begin{eqnarray}}
\newcommand{\eeq}{\end{eqnarray}}

\newcommand{\bfig}{ \left. \begin{array}{l} }
\newcommand{\bfigor}{ \left\{ \begin{array}{ll} }
\newcommand{\efig}{ \end{array} \right. }

\newcommand{\cH}{{\cal H}}

\newcommand{\e}{\epsilon}

\newcommand{\krest}{\dagger}
\newcommand{\akrest}{a^\krest}
\newcommand{\bkrest}{b^\krest}
\newcommand{\akr}{\akrest}
\newcommand{\bkr}{\bkrest}

\newcommand{\pp}{{\bf p}}

\newcommand{\ckr}{c^\krest}

\begin{document}

\title{Spin nematic states in antiferromagnets containing ferromagnetic bonds}

\author{A.\ V.\ Sizanov$^1$}
\email{alexey.sizanov@gmail.com}
\author{A.\ V.\ Syromyatnikov$^{1,2}$}
\email{syromyat@thd.pnpi.spb.ru}
\affiliation{$^1$Petersburg Nuclear Physics Institute NRC "Kurchatov Institute", Gatchina, St.\ Petersburg 188300, Russia}
\affiliation{$^2$Department of Physics, St.\ Petersburg State University, 198504 St.\ Petersburg, Russia}

\date{\today}

\begin{abstract}

The majority of recent works devoted to spin nematic phases deal with either frustrated magnets or with those described by Hamiltonians with large non-Heisenberg terms. We show in the present study that non-frustrated antiferromagnets (AFs) containing ferromagnetic (FM) bonds can show nematic phases in strong magnetic field. Among particular spin systems discussed are a ladder with FM rungs, two AF layers coupled ferromagnetically, a chain containing alternating AF and FM bonds and an AF anisotropic spin-1 chain.

\end{abstract}


\maketitle

{\it Introduction}.---Frustrated spin systems have offered in recent years a wealth of opportunities for the study of a broad range of novel types of states and phase transitions. Spin nematic phases form a class of objects in this area which has received much attention. Spin nematic states are spin-liquid-like states which show a multiple-spin ordering without the conventional long-range magnetic order. The two-spin ordering can be generally described by the tensor \cite{andreev} $Q_{jl}^{\alpha\beta} = \langle S_j^\alpha S_l^\beta\rangle-\delta_{\alpha\beta} \langle {\bf S}_j {\bf S}_l \rangle /3$. The antisymmetric part of $Q_{jl}^{\alpha\beta}$ is related to the vector chirality $\langle {\bf S}_j \times{\bf S}_l \rangle$ and describes a vector chiral spin liquid which can be stabilized in quantum spin models at $T=0$ by a sizable ring-exchange, \cite{chir} and can be found in classical frustrated spin systems at $T\ne0$. \cite{cinti,*sasha} The symmetric part of $Q_{jl}^{\alpha\beta}$ describes a quadrupolar order which has been extensively studied both theoretically and experimentally in frustrated systems with ferromagnetic (FM) and antiferromagnetic (AF) nearest-neighbor and next-nearest-neighbor couplings, respectively, in strong magnetic field $H$ (see, e.g., Ref.~\cite{syromyat} and references therein) and in magnets with large non-Heisenberg spin couplings such as biquadratic exchange $({\bf S}_1{\bf S}_2)^2$. \cite{kolezh1,*kolezh2} It has been also shown recently that quantum fluctuations accompanied by a sizable single-ion easy-axis anisotropy can also stabilize a nematic phase in the kagome spin-1 antiferromagnet. \cite{senthil} This study was motivated by recent experiments on $\rm Ni_3V_2O_8$.

It is well established that the attraction between magnons caused by frustration is the origin of quadrupolar and multipolar phases in quantum magnets. \cite{chub} In particular, the bottom of the one-magnon band lies above the lowest multi-magnon bound state at $H=H_s$, where $H_s$ is the saturation field, as a result of this attraction in magnets with FM  and AF couplings between nearest- and next-nearest neighbors, respectively. Then, transitions to nematic phases at $H<H_s$ in such systems are characterized by a softening of the multi-magnon bound-state spectrum rather than the one-magnon spectrum. 

\begin{figure}
\includegraphics[scale=0.65]{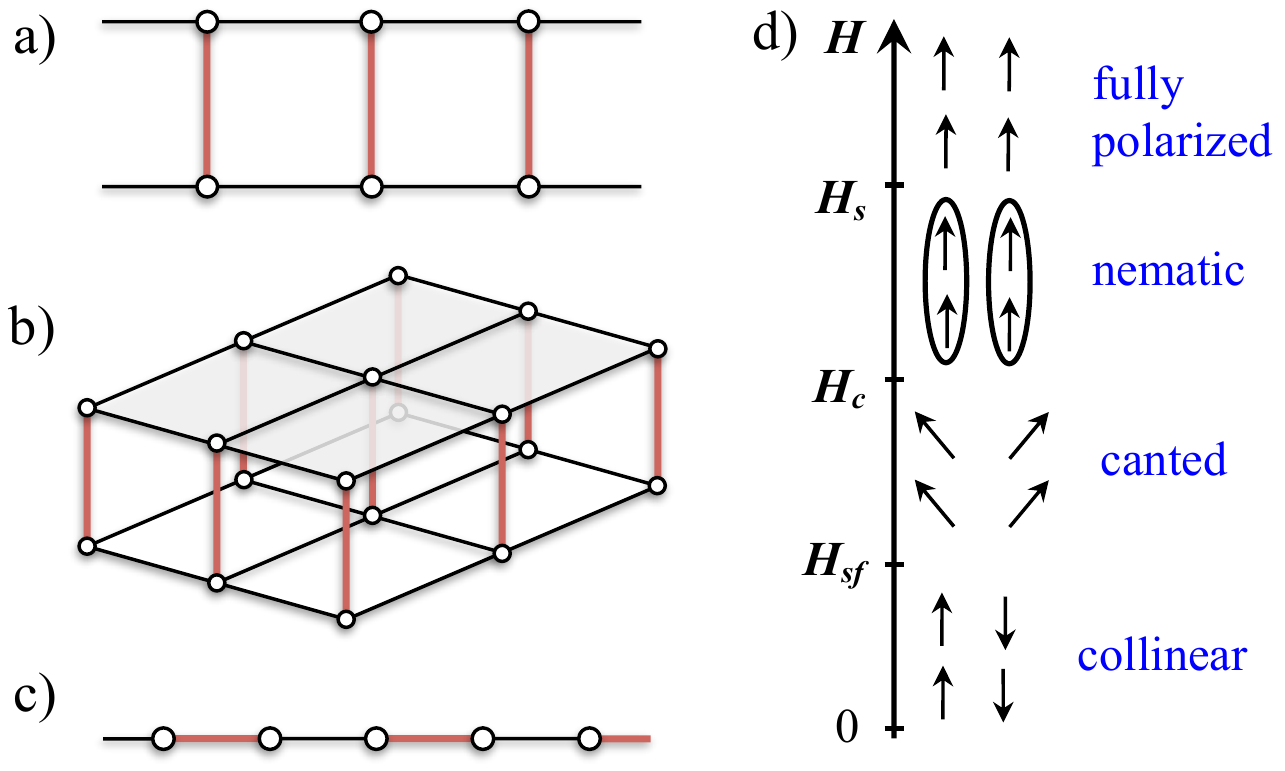}
\caption{(Color online.) (a)--(c) Spin-$\frac12$ systems which show a nematic phase in strong magnetic field and which are discussed in the present paper in some detail. Antiferromagnetic and ferromagnetic bonds are shown in black and blue, respectively. (d) A sketch of the phase diagram at small $T$ of the considered systems. The canted or the nematic phase is absent in some cases (see the text). A small inter-chain, inter-bilayer and inter-ladder couplings are taken into account.}
\label{systems}
\end{figure}

We show in the present paper that FM bonds in AF non-frustrated spin systems can also lead to the magnon attraction and to stabilization of nematic phases in strong magnetic field. To be specific, we discuss spin-$\frac12$ ladder with FM rungs, two AF layers coupled ferromagnetically and a chain containing alternating AF and FM bonds which are presented in Fig.~\ref{systems}(a)--(c). The spin ladder with FM rungs and the alternating chain have received considerable interest recently which has not been related, however, to their nematic behavior in magnetic field (see Refs.~ \cite{alm,*wat,*tot,*lang,*aris1,*aris2} and Refs.~\cite{gong,*lu,*gu,*van,*hagi}, respectively, and references therein). Although interaction between spins is antiferromagnetic in the majority of compounds containing weakly coupled spin chains, ladders or bilayers, some (mainly organic-based) materials containing FM bonds have been synthesized recently \cite{hos,*zhel} that has stimulated the theoretical activity in this field.

{\it Model and technique}.---All spin-$\frac12$ systems under discussion are described by the Hamiltonian
\begin{eqnarray}
\label{ham}
{\cal H} &=& J\sum_{\langle i,l\rangle}
\left({\bf S}_{1,i} {\bf S}_{1,l} + {\bf S}_{2,i} {\bf S}_{2,l}\right)
	- j \sum_i {\bf S}_{1,i} {\bf S}_{2,i}
	\nonumber\\
	&&{}
	- A \sum_i S^z_{1,i} S^z_{2,i}
	- H\sum_i \left(S^z_{1,i} + S^z_{2,i}\right) + {\cal H}',
\end{eqnarray}
where ${\bf S}_{n,i}$ is the $n$-th spin in the $i$-th FM bond, $j$ and $A$ are values of the isotropic and anisotropic parts of the FM exchange coupling, respectively, $\langle i,l\rangle$ denote nearest neighbor spins coupled antiferromagnetically, we set the AF coupling constant $J$ to be equal to unity in our calculations, and ${\cal H}'$ describes small inter-ladder (inter-bilayer or inter-chain) interaction.

The ground state of Hamiltonian \eqref{ham} has a collinear antiferromagnetic spin structure at small $H$ and the magnon spectrum has a gap induced by the easy-axis anisotropy $A$. There is a sequence of phase transitions upon the field increasing. 

If the easy-axis anisotropy $A$ is smaller than a critical value $A_c$ (the classical value of $A_c$ is $J_{\bf 0}$, where $J_{\bf p} = J\sum_j \exp(i{\bf p R}_{jl})$, that does not depend on $j$), the first phase transition is the first-order spin-flop one that happens when the field reaches a value $H_{sf}$. One has for the classical value of $H_{sf}$ 
\begin{equation}
\label{hsf}
H_{sf}^{cl} = \frac12\sqrt{A(2J_{\bf 0}-A)}.
\end{equation}
There is a canted AF spin structure at $H>H_{sf}$ and one of the magnon branches is gapless (as a consequence of the continuous $SO(2)$ symmetry breakdown in this phase). Our experience suggests that there would be only one phase transition to the fully polarized phase at $H=H_s$ upon further field increasing. However, we show below that in a range of parameters the fully polarized phase is preceded by a nematic one with the order parameter $\langle S^-_{1,i} S^-_{2,i}\rangle$ (see Fig.~\ref{systems}(d)). Consideration of the nematic order parameter symmetry shows that $SO(2)/Z_2$ symmetry is broken in the nematic phase (see, e.g., Ref.~\cite{syromyat}). Then, there is a Goldstone mode in the nematic phase and the transition from the canted AF phase to the nematic one is apparently of the 2D Ising type because the $Z_2$ subgroup breaks down.

If $A$ is large enough (e.g., at $A\gg J,j$), there is no canted phase. We show below that even in this case the collinear and the fully polarized gapped phases are separated by the gapless nematic one.

We examine in the present paper the possibility of the nematic phase formation just below the saturation field $H_s$ by considering the transition from the fully polarized phase. We use for this purpose the approach suggested recently by one of us in Ref.~\cite{syromyat} for high-field nematic phases analysis that is based on the following bond-operator spin representation:
\begin{align}
\label{trans}
S_{1,i}^+ =& b_i^\dagger a_i + c_i, & S_{2,i}^+ =& c_i^\dagger a_i + b_i,\nonumber\\
S_{1,i}^- =& a_i^\dagger b_i + c_i^\dagger, & S_{2,i}^- =& a_i^\dagger c_i + b_i^\dagger,\\
S_{1,i}^z =& \frac12 -a_i^\dagger a_i - c_i^\dagger c_i, & S_{2,i}^z =& \frac12 -a_i^\dagger a_i - b_i^\dagger b_i,\nonumber
\end{align}
where $a_i^\dagger$, $b_i^\dagger$ and $c_i^\dagger$ are Bose-operators which create three spin states from the vacuum $|0\rangle=\left|\uparrow\uparrow\right\rangle$ as follows: 
$a_i^\dagger |0\rangle = \left|\downarrow\downarrow\right\rangle$,
$b_i^\dagger |0\rangle = \left|\uparrow\downarrow\right\rangle$, and
$c_i^\dagger |0\rangle = \left|\downarrow\uparrow\right\rangle$,
where all spins have the maximum projection on the field direction at the state $|0\rangle$. To avoid contribution of unphysical states containing more than one particle $a$, $b$ or $c$ on a FM bond, we add to the Hamiltonian constraint terms describing an infinite repulsion between particles on each FM bond: 
$
U \sum_i ( \akr_i \akr_i a_i a_i + \bkr_i \bkr_i b_i b_i + \ckr_i \ckr_i c_i c_i + \akr_i \bkr_i a_i b_i + \bkr_i \ckr_i b_i c_i + \akr_i \ckr_i a_i c_i ),
$
where $U\to+\infty$. In particular, it is shown in Ref.~\cite{syromyat} that (i) this approach is quite convenient for discussion of the quantum phase transition to the nematic phase, (ii) along with some new results it yields also those obtained by other methods, \cite{chub,zhito,*3d,*kuzian,*kuzian2,*1d3} and (iii) although it gives quantitatively correct results at $H\approx H_s$ when $H<H_s$, it works in the nematic phase qualitatively also when $H$ is not very close to $H_s$. For the sake of self-consistency, we present below some detail of this approach. We confirm below our key analytical results by numerical ones obtained using finite cluster diagonalization technique. \cite{alps2}

Substituting Eqs.~\eqref{trans} into Hamiltonian \eqref{ham}, taking into account the constraint terms and neglecting for a moment ${\cal H}'$ one obtains for the ladder and the bilayer
\beq
\label{h2}
\cH_2 &=& \sum_{\bf p} 
\left(
(\bkr_\pp b_\pp + \ckr_\pp c_\pp ) \left(H + \frac12 (J_\pp - J_{\bf 0} ) + \frac12 (j+A) \right)
\right.\nonumber\\
&&{}
\left.
-\frac12 j \left( \bkr_\pp c_\pp + \ckr_\pp b_\pp \right)  
+ \akr_\pp a_\pp ( 2H - J_{\bf 0} ) \right),  \\
\label{h3}
\cH_3 &=& \frac{1}{2\sqrt N}\sum
\left(
\akr_1 b_2 c_3 \left( J_2 + J_3 \right) + 
\bkr_1 \ckr_2 a_3 \left( J_1 + J_2 \right)\right), \\
\label{h4}
\cH_4 &=& \frac1N\sum
\left(
\akr_1 \akr_2 a_3 a_4 \left( J_{1-3} + U \right)+ U\bkr_1 \ckr_2 b_3 c_3 
\right.
\nonumber\\
&&	+\left( \akr_1 \bkr_2 a_3 b_4 + \akr_1 \ckr_2 a_3 c_4 \right) \left( J_{1-3} + \frac12 J_{1-4} + U \right)
\nonumber\\
&&{}
\left.
	+\left( \bkr_1 \bkr_2 b_3 b_4 + \ckr_1 \ckr_2 c_3 c_4 \right) \left( \frac12 J_{1-3} + U \right)
	\right), 
\eeq
where $N$ is the number of FM bonds, the momentum conservation laws $\sum_i{\bf p}_i={\bf 0}$ are implied in Eqs.~\eqref{h3} and \eqref{h4}, and we omit some indexes ${\bf p}$ in Eqs.~\eqref{h3} and \eqref{h4}. It is convenient to introduce the following Green's functions:
\begin{eqnarray}
\label{gadef}
 G_a ( k ) &=& - i \langle a_k a^\dagger_k \rangle, \\
\label{gbdef}
 G_b ( k ) &=& - i \langle b_k b^\dagger_k \rangle, \quad G_c ( k ) = - i \langle c_k c^\dagger_k \rangle, \\
 \label{fdef}
 F ( k ) &=& - i \langle b_k c^\dagger_k \rangle, \quad {\overline F} ( k ) = - i \langle c_k b^\dagger_k \rangle.
\end{eqnarray}
Poles of $G_a ( k )$ give the spectrum of $a$ particles, whereas poles of $G_b ( k )$, $G_c ( k )$, $ F ( k )$ and ${\overline F} ( k )$ which have the same denominator determine spectra of two one-magnon branches. \cite{see}

Particles $b$ and $c$ carrying spin 1 are of the one-magnon nature. Their spectra calculated at $H>H_s$ using Eqs.~\eqref{trans} coincide with those derived using the conventional approaches such as the Holstein-Primakoff transformation. It can be shown \cite{syromyat} that spectra of one-magnon excitations are determined solely by ${\cal H}_2$ (Eq.~\eqref{h2}) at $H\ge H_s$ and they have the form
\beq
\e_1 (\pp) &=& H + \frac12 \left( J_\pp - J_{\bf 0} \right) + \frac12 A,\\
\e_2 (\pp) &=& \e_1 (\pp) + j.
\eeq
The lower branch $\e_1 (\pp)$ has a minimum at $\pp=\pi$ (or $(\pi,\pi)$) and a gap which closes when $H$ becomes equal to
\begin{equation}
\label{hc}
H_c = J_{\bf 0} - A/2.
\end{equation}

\begin{figure} 
\includegraphics[scale=0.52]{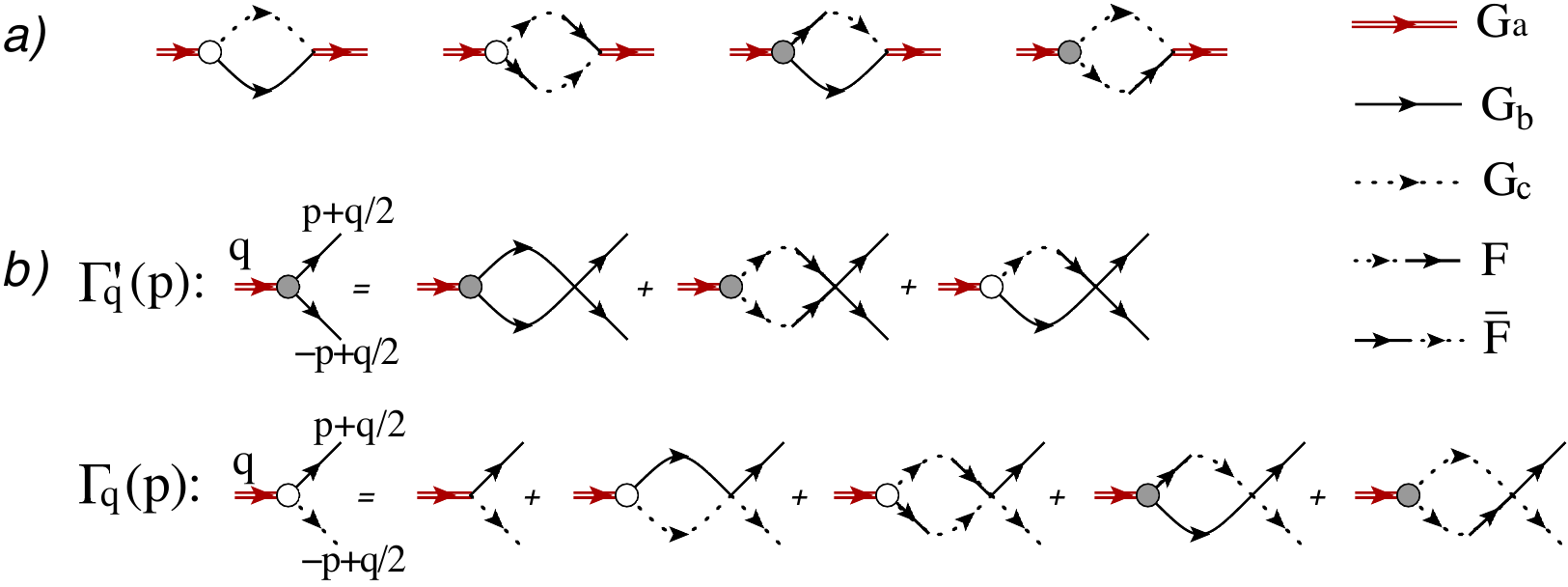}
\caption{(Color online.) a) Diagrams for the self-energy part $\Sigma_a(k)$ of $a$-particles at $H\ge H_s$. Green's functions of $b$- and $c$- particles $G_b(p)$, $G_c(p)$, $F(p)$ and $\overline{F} ( p )$ are defined by Eqs.~\eqref{gbdef}--\eqref{fdef}. Circles represent renormalized vertices. Equations for the vertices are presented in panel (b). Bare vertices are defined by Eqs.~\eqref{h3} and \eqref{h4}.
\label{diag1}}
\end{figure}

In contrast to $b$ and $c$ particles, $a$ particles carrying spin 2 are of the two-magnon nature. Their spectrum coincides with the two-magnon bound-state spectrum found using conventional methods. \cite{see} To find the spectrum $\e_a(\pp)$ of $a$ particles at $H\ge H_s$ one has to take into account diagrams shown in Fig.~\ref{diag1}(a) which contain three-particle vertexes. Equations for them are presented in Fig.~\ref{diag1}(b). The minimum of $\e_a(\pp)$ is at $\pp={\bf 0}$ and the gap in $\e_a(\pp)$ closes at $H=H_c'$. If $H_c>H_c'$, the transition takes place at $H=H_c=H_s$ to the canted phase which can be described as the Bose-Einstein condensation (BEC) of one-magnon excitations. As it is seen from Eqs.~\eqref{trans}, $\langle{\bf S}^\perp_{n,i}\rangle$ becomes finite in this case, where $\perp$ denotes the projection on the $xy$ plane. In contrast, if $H_c<H_c'$ the transition from the fully polarized phase to the nematic one takes place at $H=H_c'=H_s$ which can be described within our approach as the BEC of $a$ particles. It is seen from Eqs.~\eqref{trans} that $\langle{\bf S}^\perp_{n,i}\rangle=0$ in the nematic phase and $\langle S^-_{1,j} S^-_{2,j}\rangle\equiv\langle a_j^\dagger\rangle\propto e^{-i\phi}\sqrt{H_s(T)-H}$ is the nematic order parameter (for ${\cal H}'\ne0$), where $\phi$ is an arbitrary phase. \cite{syromyat} All the static two-spin correlators decay exponentially in the nematic phase. We calculate below $\e_a(\pp)$ for the selected spin systems and find stability conditions of the nematic states.

{\it AF ladder with FM rungs (see Fig.~\ref{systems}(a)).}---Unfortunately, the general expression for $\e_a(\pp)$ is quite complicated but it is simplified greatly in the limiting case of the Ising exchange on FM bond (i.e., at $j=0$). To illustrate main properties of the nematic phase we consider here in some detail the case of $j=0$. In particular, one finds at ${\cal H}'=0$ and $H\ge H_s$
\beq
\label{eal}
\e_a(\pp) = 2H - 2 + A - \sqrt{4\cos^2 \frac{p}{2} + A^2}.
\eeq
The gap in $\e_a(\pp)$ vanishes at $H_s=1+\sqrt{1+A^2/4}-A/2$ that is larger than $H_c=2-A/2$ (see Eq.~\eqref{hc}) at all positive $A$. Then, the nematic phase is stable at $A>0$ when $j=0$. Interestingly, the nematic phase arises also in the case of $A\gg1$, when the canted phase is absent. Thus, the gapless nematic phase separates two gapped phases at large $A$, the fully polarized and the collinear ones. 

Spectrum \eqref{eal} has a quadratic dispersion near its minimum at $p=0$: $\e_a(\pp) \approx 2H - 2 + A - \sqrt{4 + A^2}+D_\|p^2$, where $D_\|=1/2\sqrt{4 + A^2}$. The spectrum remains quadratic also after taking into account a small ${\cal H}'$ that leads to the 3D BEC relation $H_s(0)-H_s(T)\propto T^{3/2}$. \cite{syromyat}

One obtains in the first order in $\rho$ and in the leading order in the inter-ladder interaction as it is done in Ref.~\cite{syromyat} for quasi-1D frustrated systems
\begin{eqnarray}
\label{mmcor}
&&\left\langle S_{1,j}^-S_{2,j+n}^- \right\rangle = \sqrt\rho e^{-i\phi} \frac{4}{A^2}\left(\frac A2-\sqrt{1+\frac{A^2}{4}}\right)^n,\\
\label{pmcor}
&&\left\langle S_{1,j}^+S_{1,j+n}^- \right\rangle = \left\langle S_{2,j}^+S_{2,j+n}^- \right\rangle  \\
&&=\rho \frac{4}{A^2}\left( \frac{4n}{A^2} + \frac{4\sqrt{4+A^2}}{A^3} -2 \right)
\left(\frac A2-\sqrt{1+\frac{A^2}{4}}\right)^n,\nonumber\\
\label{zzcor}
&&\left\langle \left( S_{1,j}^z - \frac12 \right) \left( S_{2,j+n}^z- \frac12\right) \right\rangle = 
\left|\left\langle S_{1,j}^-S_{2,j+n}^- \right\rangle\right|^2,
\end{eqnarray}
where $n>0$ and $\rho=\langle a_i^\dagger a_i\rangle\propto(H_s(T)-H)$ is the "condensate" density. The rest two-spin static correlators containing $S^+$ or $S^-$ are equal to zero and $\langle S_{1,j}^z S_{1,j+n}^z\rangle = \langle S_{2,j}^z S_{2,j+n}^z\rangle\sim\rho^2$. It is seen from Eqs.~\eqref{mmcor}--\eqref{zzcor} that all correlators decay exponentially at $A>0$.

One obtains for the magnetization
\begin{equation}
\label{mag}
	\langle S^z_{q,j} \rangle = \frac12 - \left( 2 - \frac{8}{A^2} + \frac{16\sqrt{4+A^2}}{A^5} \right) \rho,
\end{equation}
where $q=1,2$. It may seem that Eqs.~\eqref{mmcor}--\eqref{mag} are invalid for arbitrary small $A$. However, the above results are valid in the near vicinity of $H_s$ (i.e., at very small $\rho$) in this case because $H_s$ and $H_c$ merge in the limit $A\to0$.

It is implied in Eqs.~\eqref{mmcor}--\eqref{mag} that ${\cal H}'\ne0$ so that $\rho\ne0$. The situation is completely different in the purely 1D case because $\rho\equiv0$ and there is no long range nematic order. Bearing in mind the quadratic dispersion of $\e_a(\pp)$ near its minimum and using results of 1D Bose-gas discussions \cite{lieb,*lieb2,*kor} one obtains \cite{see} at $H\approx H_s$ and $T=0$
\begin{eqnarray}
\label{mag1d}
	&&\frac12 -  \left\langle S_{q,j}^z\right\rangle = \langle a_j^\dagger a_j \rangle = \frac1\pi \sqrt{\frac{2(H_s-H)}{D_\|}},\\
\label{zz}
&&\left\langle S_{q,j+n}^z(t) S_{p,j}^z(0) \right\rangle 
\approx 
\left\langle S_{j}^z \right\rangle^2 
\\
&&- \frac{1}{\pi}\left(\frac{1}{(n+iut)^2} + \frac{1}{(n-iut)^2}\right) 
+ B_1\frac{\cos(\pi\langle a_j^\dagger a_j \rangle n)}{n^2+u^2t^2},\nonumber\\
\label{nemcorr}
&&\left\langle S_{1,0}^+(t)S_{2,0}^+(t) S_{1,n}^-(0) S_{2,n}^-(0) \right\rangle 
\approx \frac{B_2}{\sqrt{|n+iut|}},
\end{eqnarray}
where $q,p=1,2$, $n\to\infty$, $u=4\pi D_\|\langle a_j^\dagger a_j \rangle$ and $B_{1,2}$ are constants.

\begin{figure} 
\includegraphics[scale=0.25]{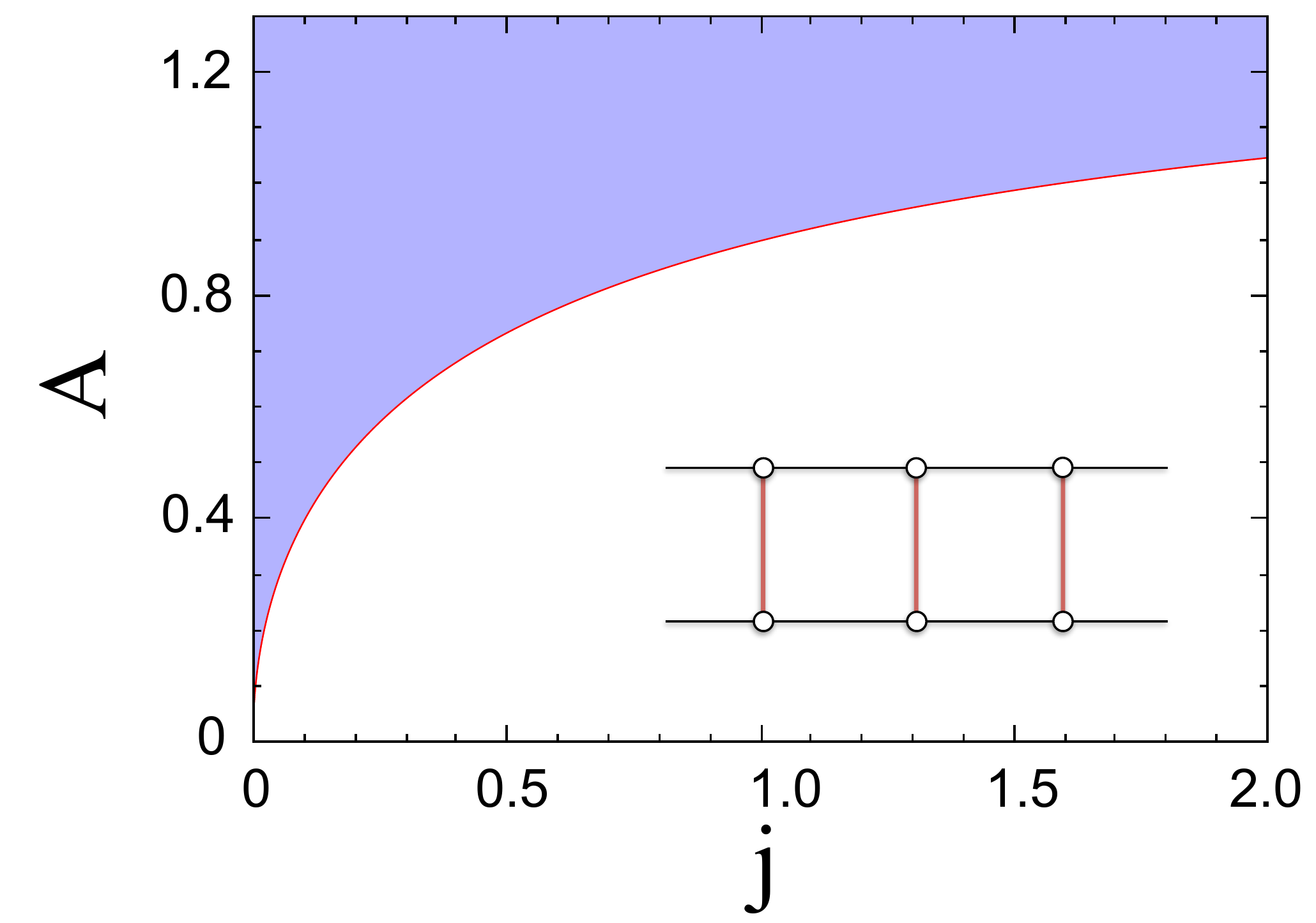}
\includegraphics[scale=0.25]{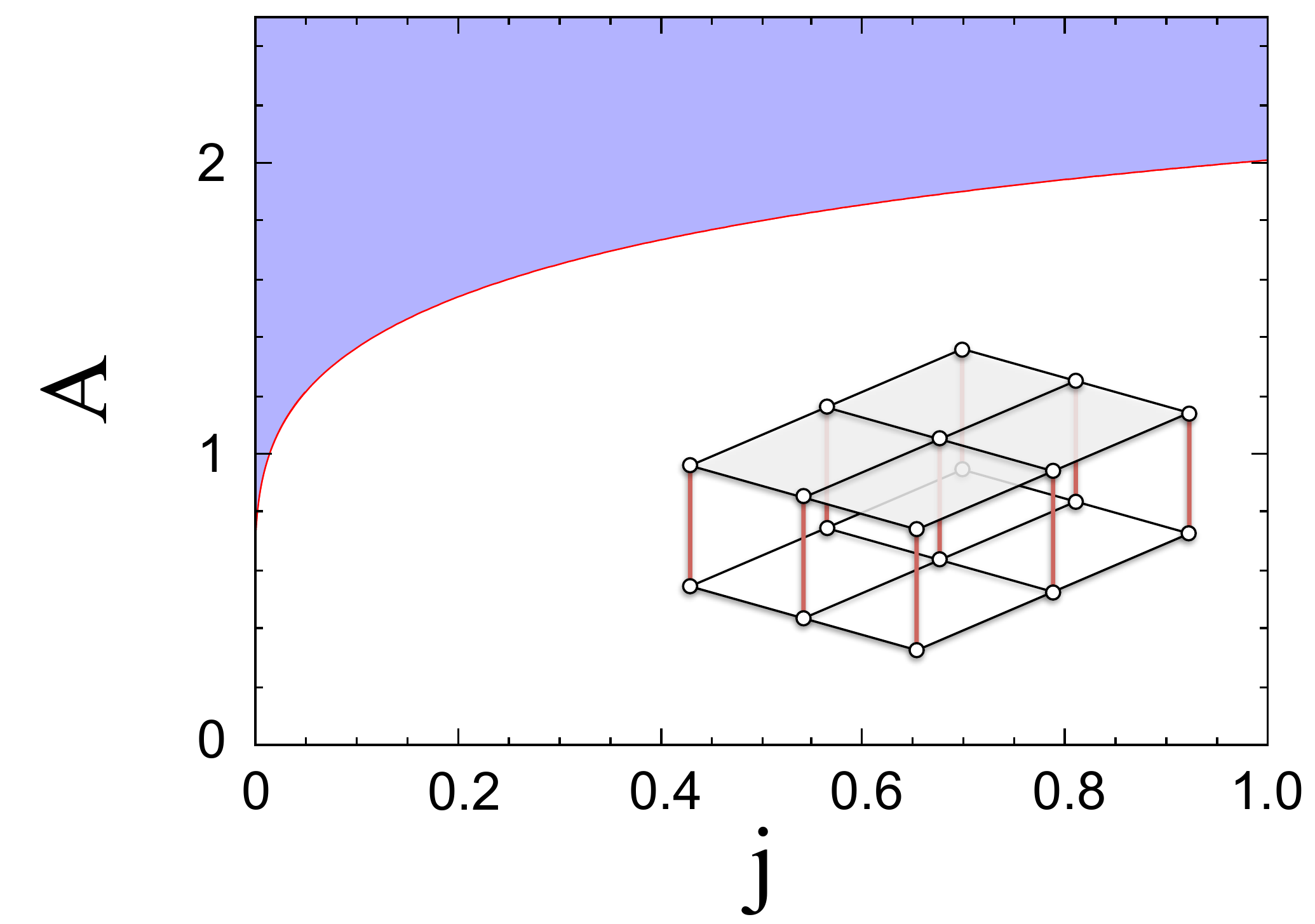}
\includegraphics[scale=0.25]{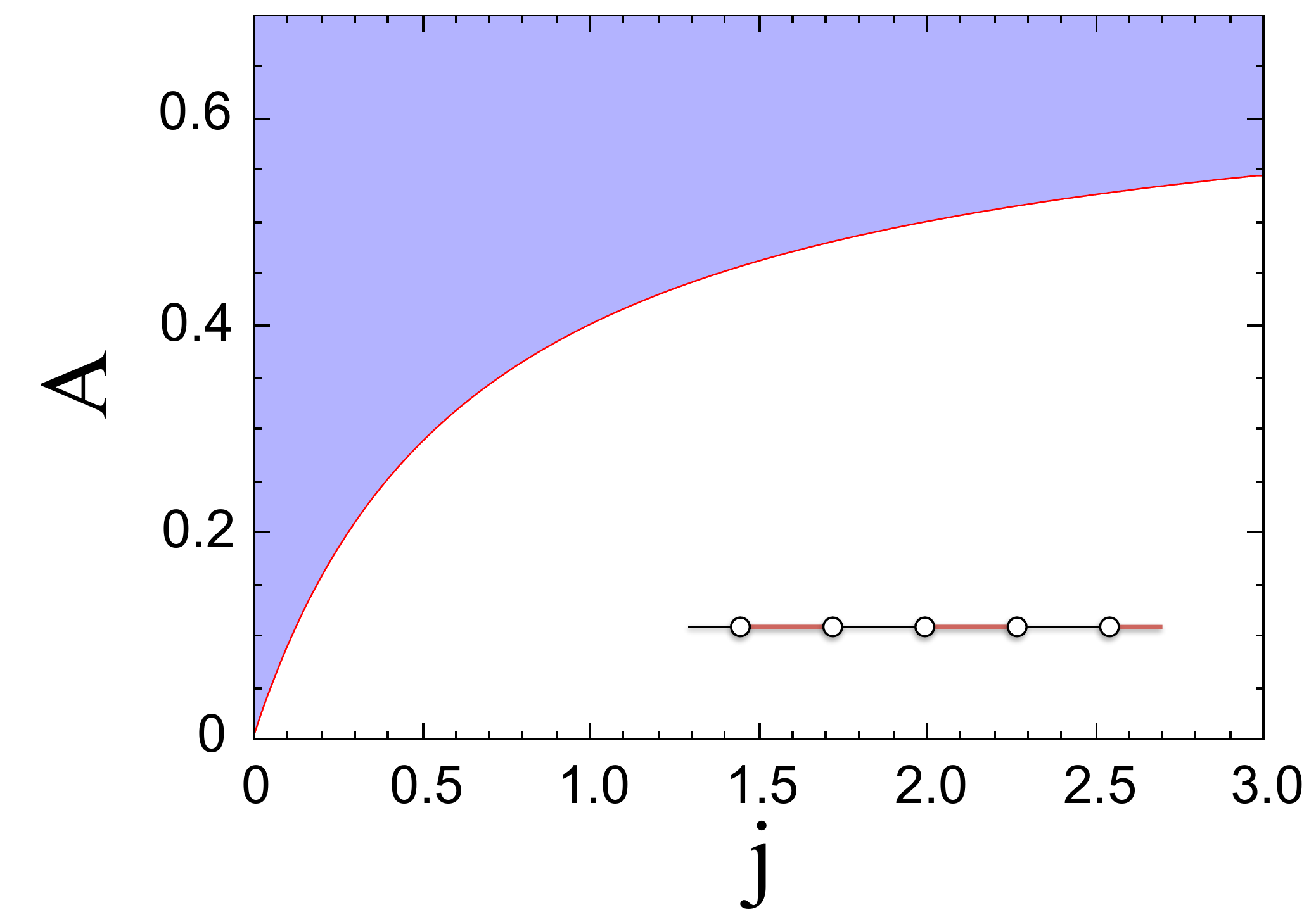}
\caption{(Color online.) Regions are shown of the nematic phase stability in the ladder with FM rungs (inequality \eqref{inlad}), in the bilayer with FM coupling (inequality \eqref{inbi}), and in the alternating chain.
\label{stabfig}}
\end{figure}

Analysis of the general expression for $\e_a(\pp)$ gives the following general criterion of the nematic phase stability:
\beq
\label{inlad}
A > \frac{ 4 j }{ j + 2 \sqrt{ j ( j + 2) } }
\eeq
that is shown graphically in Fig.~\ref{stabfig}. One concludes from Eq.~\eqref{inlad} that only large $A\agt j$ can stabilize the nematic phase if $j\sim1$. In contrast, quite small anisotropy on FM bonds $j\gg A\sim1$ is sufficient for this purpose if $j\gg1$.

In the limiting case of $j\to\infty$, our model describes the spin-1 AF chain with the easy-axis single-ion anisotropy $A$. The nematic order parameter reads in this case as $\langle (S_i^-)^2\rangle$ and one obtains from Eq.~\eqref{inlad} that the nematic phase is stable at $A>4/3$. Similar nematic phase is discussed in Ref.~\cite{senthil} in spin-1 kagome AF with large single-ion easy-axis anisotropy at $H=0$.

Our finite cluster calculations confirm that the transition takes place at $H=H_s$ to the nematic phase when inequality \eqref{inlad} holds. This numerical consideration is simplified by the fact that the Hamiltonian \eqref{ham} commutes with the $z$ component of the total spin ${\cal S}^z$ and with the Zeeman term. As a result all the Hamiltonian eigenstates can be classified by eigenvalues $M$ of ${\cal S}^z$. Let us denote $E(M)$ the minimum energy in each $M$ sector at $H=0$. The ground state energy of a cluster with $L$ spins in magnetic field is given by the minimum value of $E(M)-HM$. An important observation is that values $E(M)-HM$ at even $M_{sat}-M$ are smaller than those with odd $M_{sat}-M$ when $H$ is close to its saturation value, where $M_{sat}=L/2$. Thus, one can expect that a condensation takes place in the thermodynamic limit of elementary excitations carrying spin 2. Then, the lowest state in each even-$(M_{sat}-M)$ sector has zero momentum (if the periodic boundary conditions are applied) that is also in agreement with the bound states condensation scenario. Values of $H_s$ obtained numerically as a result of analysis of clusters with $L=8\div48$ are in excellent agreement with the corresponding values found analytically. Numerical consideration of clusters with $L=8\div26$ similar to that performed in Refs.~\cite{sakai1,*sakai2} confirms also the validity of Eq.~\eqref{mag1d}. For example, one obtains for $j=0$ and $A=2$ that $\frac12 - \left\langle S_{q,j}^z\right\rangle = \alpha(H_s-H)^{1/\delta}$, where $\alpha=1.04\pm0.11$ and $\delta=2.05\pm0.05$, that is in good agreement with the analytical result $\langle a_j^\dagger a_j \rangle \approx 1.07\sqrt{H_s-H}$.

{\it AF bilayer with FM coupling (see Fig.~\ref{systems}(b)).}---The analytical analysis of this system is much more complicated than that carried out above for the ladder. Then, we restrict ourselves by discussion of the region of the nematic phase stability that is defined by inequality (see Fig.~\ref{stabfig})
\beq
\label{inbi}
&&A > \frac{ 8  - 8j f(j) }{ ( 8 + j ) f(j) -1  }, \\
&&f(j) = \frac{1}{(2\pi)^2} \int_0^{2\pi} \frac{dxdy}{ j + 2 + \cos x + \cos y}.
\eeq
One concludes from Eq.~\eqref{inbi} that, similar to the ladder with FM rungs, only large $A\agt j$ can stabilize the nematic phase if $j\sim1$. In contrast, quite small anisotropy on FM bonds $j\gg A\sim1$ is sufficient for this purpose if $j\gg1$.

{\it Alternating chain. (see Fig.~\ref{systems}(c))}---Particular expressions are quite cumbersome in this case. Then, we restrict ourselves by graphical representation of the nematic phase stability region which is also shown in Fig.~\ref{stabfig}. It is seen that the graphic resembles those for the ladder and the bilayer. Numerical consideration confirms existence of the nematic phase and the validity of Eq.~\eqref{mag1d}. Numerical values of $H_s$ are in excellent agreement with analytical results.

{\it To conclude}, we demonstrate that FM bonds in non-frustrated antiferromagnets can lead to nematic spin states in strong magnetic field. The uniaxial anisotropy on FM bond is necessary for the nematic phase stabilization in all the considered systems. The present study should stimulate further theoretical and experimental activity in nematic phases discussion both in the considered systems and in other ones containing FM bonds.

This work was supported by the President of the Russian Federation (Grant No.\ MD-274.2012.2), the Dynasty foundation and RFBR Grants No.\ 12-02-01234 and No.\ 12-02-00498.

\bibliography{nemlib} 

\end{document}